\documentclass[12pt]{article}
\def \g {\gamma}
\def \L {\Lambda}

\newcommand{\rf}[1]{(\ref{#1})}
\def \ci {\cite}
\def \del{\partial}

\def \la {\label}
\def \bi {\bibitem}
\def \G {\Gamma}
\def \const{{\rm const}}
\def \vt {\vartheta}
\def \S {{\cal K}}

\def \ads {$AdS_5 \times S^5$ }
\def \ov {\over}

\def \k {\kappa}

\def \ha {{1 \ov 2}}

\def \ep {\epsilon}

\newcommand{\be}{\begin{equation}}
\newcommand{\ee}{\end{equation}}
\newcommand{\bea}{\begin{eqnarray}}
\newcommand{\eea}{\end{eqnarray}}
\newcommand{\eps}{\epsilon}

\newcommand{\Pp}{{\cal P}_+}
\newcommand{\Pm}{{\cal P}_-}
\newcommand{\Ppm}{{\cal P}_\pm}

\newcommand{\s}{\sigma}

\newcommand{\ksym}{$\kappa$-symmetry }

\begin{document}
\vspace*{-.6in}
\thispagestyle{empty}
\begin{flushright}
SU-ITP-99/4\\
hep-th/9901095
\end{flushright}
\baselineskip = 20pt

\vspace{.4in}
{\Large \bf
\begin{center}
Black Holes, Branes and\\

\

Superconformal Symmetry\footnote{Work
supported by the NSF
grant PHY-9870115.}
\end{center}}

\vskip 0.7 cm

\begin{center}
{\bf Renata Kallosh}\\
\emph{Physics Department,  Stanford University, Stanford, CA  
94305-4060, USA}
\end{center}
\vspace{0.7 cm }

\begin{center}
\textbf{Abstract}
\end{center}
\begin{quotation}
\noindent
The main focus of this lecture is on extended objects in $adS_{p+2}\times
S^{d-p-2}$ bosonic backgrounds with unbroken supersymmetry. The  
backgrounds are
argued to be exact,  special consideration are given to the non-maximal
supersymmetry case. The near horizon superspace construction is  
explained. The
superconformal symmetry appears in the worldvolume actions as the  
superisometry
of the near horizon superspace, like  the superPoincar\'{e} symmetry  
of  GS
superstring and BST supermembrane in the flat superspace. The issues  
in  gauge
fixing of local kappa-symmetry are reviewed.

We describe the features of the gauge-fixed IIB superstring in  
$adS_{5}\times
S^{5}$ background with RR 5-form. From a truncated boundary version  
of it we
derive an analytic N=2 off shell harmonic superspace of Yang-Mills  
theory. The
reality condition of the analytic subspace, which includes the  
antipodal map on
the sphere, has a simple meaning of the symmetry of the string  
action in the
curved space. The relevant issues of black holes and superconformal  
mechanics
are  addressed.

\end{quotation}

\vfil
\centerline{To be published in the Proceedings of}
\centerline{\it Quantum Aspects of Gauge Theories, Supersymmetry, and
Unification}
\centerline{Corfu, Greece -- September 1998}

\newpage

\pagenumbering{arabic}


During the last year there was some progress in establishing connections
between exact solutions of the supergravity,  near horizon geometry  
of black
holes and branes, and quantum field theories with (super)conformal  
symmetries.
The connection relies on the (super)isometries
of the configurations which are products of anti de Sitter space and  
a sphere,
$adS*S$ and which also are characterized  by some charge since the
configurations have a non-trivial form field. The (super)isometries  
of these
exact configurations form a superconformal algebra. The purpose of  
this lecture
is to discuss the set of connections between such configurations of  
space-time
and (super)conformal theories.

This lecture is based mostly on my recent  work with P. Claus, J.  
Kumar, A. van
Proeyen,  A. Rajaraman, J. Rahmfeld, P. Townsend,
and A. Tseytlin and on numerous discussions with my collaborators. I  
will cover
here some aspects of our work which relate the extended  
supersymmetric objects
in space-time and the worldvolume actions with superconformal symmetry.
In this lecture I rely on few other  pedagogical lectures at this  
school where
many aspects of M-theory, string theory and ADS/CFT Maldacena's duality
\cite{Malda}  were explained. The contributions to this proceedings  
by P. Claus
and P. Termonia have an overlap with this lecture and may be useful  
to read in
this context.  The topics to be covered in the lecture are:

\

\noindent 1. Black holes and branes as solutions of supergravity  
equations in
space-time.

\

\noindent 2. Is the supersymmetric $adS_{p+2}\times S^{d-p-2}$ geometry
conformally flat?

\

\noindent 3. Exactness of $adS_{p+2}\times S^{d-p-2}$ +form vacuum;  
special
consideration for  $adS_{2}\times S^{2}$ +2-form near horizon black  
hole case
with smaller  supersymmetry.

\

\noindent 4. Flat superspace background,  its isometry and  
superPoincar\'{e}
symmetry of the extended objects; near horizon superspace, its   
isometry and
superconformal symmetry of the extended objects.

\

\noindent 5. Choices of gauges for fixing $\kappa$-symmetry.

\

\noindent 6. IIB Green-Schwarz superstring in $adS_{5}\times S^{5}$  
background
with RR 5-form.

\

\noindent 7. Analytic N=2 harmonic superspace from the quantization of the
truncated GS string (superparticle)  in the curved background of   
the adS*S
boundary. The role of the antipodal map on the sphere.

\

\noindent 8. Black holes and superconformal mechanics of a particle  
approaching
the black hole horizon.

 \

An attempt will be made to describe the topics listed above in a  
relatively
simple way, mostly to explain the new concepts, referring the reader  
to the
published papers for the details.


\section{Black Holes and Branes as solutions of supergravity equations in
space-time.}

Consider various  supergravity theories in d-dimensions. Black holes  
($p=0$)
and higher branes ($p>0$) are solutions of supergravity field  
equations. The
metric has the form
\begin{equation}
ds^2_{brane}= H^{-{2\over p+1}}dx^\mu \eta_{\mu\nu} dx^\nu +  
H^{{2\omega \over
p+1}} dy^m dy^m
\end{equation}
Here $x^\mu$ are the $(p+1)$ coordinates along the brane, $y^m$ are  the
remaining $(d-p-1)$   coordinates of the space orthogonal to the  
brane, $y^m
y^m \equiv r^2$ and $H$ is a harmonic function  in $(d-p-1)$-dimensional
transverse space:
\begin{equation}
H= 1+ \left ({R\over r}\right) ^{d-p-3}
\end{equation}
Such metric has to be supplemented by some form-field $F\sim   
N\times $volume
so that the configuration has the maximal amount of unbroken  
supersymmetry.
This leads to the interpretation of this solution as a set of  $N$  
parallel
branes on top of each other. The number $N$ is proportional to $R$ in some
power, where $R$ is the parameter in the harmonic function $H$.
When the parameter $\omega$ picks up the  `magic' value \cite{conf}
\begin{equation}
\omega={p+1\over d-p-3} \ ,
\label{omega}\end{equation}
the metric given above becomes a metric with the non-singular near horizon
geometry at $r=0$. The limiting metric at either very large $N$ when
$R\rightarrow \infty$ or near the horizon at $r\rightarrow 0$ is
\begin{equation}
ds^2_{nearhor}=  \left ({r\over R}\right) ^{2 \omega}dx^\mu \eta_{\mu\nu}
dx^\nu +  \left ({R\over r}\right) ^2 dy^m dy^m
\label{cartesian}\end{equation}
This metric is not yet in a form of the product space $adS_{p+2}\times
S^{d-p-2}$ since here the cartesian coordinates of the d-dimensional  
target
space are used and the split of $d$-dimensions into $p+1$ and $d-p-1$ is
performed. To see that this metric is actually the $adS_{p+2}\times  
S^{d-p-2}$
one has to switch to spherical coordinates of the transverse space
$
d\vec y^2= dr^2 + r^2 d^2\Omega
$,
which gives
\begin{equation}
ds^2_{adS*S}= ds^2_{adS} +ds^2_{S} =  \left ({r\over R}\right) ^{2
\omega}dx^\mu \eta_{\mu\nu} dx^\nu +  \left ({R\over r}\right) ^2  
dr^2 +  R^2
d^2 \Omega
\label{spherical}\end{equation}
Here the first two terms give the metric of $adS_{p+2}$ space $ds^2_{adS}=
\left ({r\over R}\right) ^{2 \omega}dx^\mu \eta_{\mu\nu} dx^\nu +  \left
({R\over r}\right) ^2 dr^2$  and the third term gives the metric of  
$S^{d-p-2}$
sphere  $ds^2_{S}=R^2 d^2 \Omega $. Now we have a split of the  
original $d$
dimensions into $p+2$ and $d-p-2$ coordinates of the product space.

The advantage of using {\it cartesian coordinates} is that the  
$R$-symmetry of
the superconformal algebra, $SO(d-p-1)$ is manifest. In this  
coordinate system
also the action of GS superstring in $adS_5\times S^5$ background is the
simplest.

The advantage of using the {\it spherical coordinates} is that the  
supercoset
construction is simple.

It is important to realize that what in space-time is a {\it  
coordinate which
labels the points in space-time}, in the worldvolume actions becomes  
a field,
depending  on the worldvolume coordinates. This means that the  
properties of
coordinate system in space-time transfer into choice of the {\it  
coordinate
system in the space of fields} on the worldvolume which may lead to  
various
possibilities to develop a quantum theory.

The metric of $adS_{p+2}\times S^{d-p-2}$ with the form $F$ has an  
enhancement
of unbroken supersymmetries comparative to the full brane metric.  
This means
that the Killing spinor equation for the supersymmetry  
transformation rules of
gravitino (and dilatino) has a solution with the maximal amount of  
the zero
modes.
\begin{equation}
\delta \psi (g, F) =\nabla \epsilon + F \epsilon =0 \ ,  \qquad  
\epsilon \neq 0
\label{kil}\end{equation}


\section{Is  the supersymmetric $adS_{p+2}\times S^{d-p-2}$ geometry
conformally flat?
}
{}From questions I had during the school it become clear to me that  
there is a
confusion with respect to the issue of conformal flatness of  
$adS_{p+2}\times
S^{d-p-2}$ geometries. To enhance this confusion and explain its  
source I will
bring up here the private statement of S. Hawking and G. Horowitz  
who observed
that this geometry is conformally flat for all cases contrary to the  
claim in
\cite{conf}.  The resolution of this controversy is the following.
If the issue of supersymmetry is ignored, the metric of
$adS_{p+2}\times S^{d-p-2}$  geometry can be taken in the form
\begin{equation}
ds^2_{adS*S}= ds^2_{adS} +ds^2_{S} =  \left ({r\over R}\right) ^{2 }dx^\mu
\eta_{\mu\nu} dx^\nu +  \left ({R\over r}\right) ^2 dr^2 +  R^2 d^2 \Omega
\end{equation}
We will show below that this metric is conformally flat. However if  
the metric
$g$ with the form $F$ are required to solve the Killing spinor equation
(\ref{kil}) in addition to solving the field equations, the choice of the
parameter $\omega$ is not arbitrary and depends on $d$ and $p$ as  
shown in eq.
(\ref{omega}).
Thus for the supersymmetric  solution given in the previous section  
only for
\begin{equation}
\omega={p+1\over d-p-3}=1 \ ,
\end{equation}
we have a conformally flat metric.
For example,  for D3 brane, $d=10,  p=3$, a self-dual string  $d=6,  
p=1$ and
for black holes,  $d=4, p=0$ the conformal flatness takes place even in
supersymmetric case.
However, for
\begin{equation}
\omega={p+1\over d-p-3}\neq 1 \ ,
\end{equation}
in cases like M2 brane $d=11, p=2$, M5 brane $d=11, p=5$, for black  
holes $d=5,
p=0$, for magnetic string $d=5, p=1$ the  metric of the  
configuration with the
unbroken supersymmetry is not conformally flat!

To understand it better let us perform a change of coordinates
\begin{equation}
 \left ({R\over r}\right)= z^\omega
\end{equation}
The metric (\ref{spherical}) of the supersymmetric solution becomes
\begin{equation}
ds^2_{adS*S} = {1\over z^2} dx^\mu \eta_{\mu\nu} dx^\nu + (\omega R)  
^2 {dz^2
\over z^2} +  R^2 d^2 \Omega
\end{equation}
Now we rescale $x= \omega R  \tilde x$ and the metric becomes
\begin{equation}
ds^2_{adS*S} = {(\omega R)^2\over z^2} \left [d\tilde x^\mu \eta_{\mu\nu}
d\tilde x^\nu  + dz^2 \right ] +  R^2 d^2 \Omega
\end{equation}
One more step is required to combine the angles of the sphere with  
the radial
direction $z$ into $d-p-1$ coordinates $z^m$ so that
$
d\vec z^2 = dz^2  +  z^2 d^2 \Omega
$.
This step is possible only for $\omega=1$, the basic reason being  
the fact that
{\it one can not rescale the angles of the sphere} and therefore
\begin{equation}
{(\omega R)^2\over z^2} dz^2  + z^2 R^2 d^2 \Omega
\end{equation}

In cases when $\omega =1$ which also means that the dimension of the  
adS space
equals the dimension of the sphere,
\begin{equation}
 p+2= d-p-2 \ ,
\end{equation}
we have found that the metric $adS_{p+2}\times S^{p+2}$  of the  
supersymmetric
configuration  is conformally  flat  in the full target space $d$
\begin{equation}
ds^2_{adS_n *S^n} = { R^2\over z^2} \left [d  \vec x^2 + d\vec z^2 \right
]\end{equation}
In case of $adS_5\times S^5$,  $adS_3\times S^3$ and $adS_2\times S^2$ the
metric of the supersymmetric near horizon configuration is  
conformally flat in
d=10 for D3 brane, in d=6 for the string and and in d=4 for black holes
respectively.

The metric of the supersymmetric near horizon configuration  
$adS_4\times S^7$
of the M2 brane and of $adS_7\times S^4$ of the M5 brane in d=11  
(and other
cases with $p+2\neq d-p-2$) is not conformally flat!

Having removed the confusion about the conformal  
flatness/non-flatness of the
generic supersymmetric $adS_{p+2}\times S^{d-p-2}$ solution of classical
supergravity equations, we may address another related controversial issue
concerning the exactness of such configurations in the framework of the
effective action of supergravities with all higher derivative terms  
present.

The existing lore relates the absence of quantum corrections to conformal
flatness of the metric. Partially this is based on the proof  
presented by Banks
and Green  \cite{BG} that in d=10 IIB supergravity $(R_{abcd})^4$  
terms do not
correct the $adS_5\times S^5$ metric since these terms actually  
depend on the
Weyl tensor $C_{abcd}$ which vanishes for the conformally flat metric.

Now that we have clearly shown that for M2 and M5 brane there is no  
conformal
flatness of the supersymmetric metric, we will explain how the exactness
argument works and why in the particular case of $(R_{abcd})^4$  
terms in IIB
supergravity this more general argument is reduced to conformal flatness.


\section{Exactness of $adS_{p+2}\times S^{d-p-2}$ +form vacuum
}

We have argued so far that $adS*S$ spaces with form-fields are  
solutions of
classical equations of supergravities. Suppose that we have an  
effective action
of supergravity where all possible terms with higher derivatives  
compatible
with supersymmetries are added to the action. We know their structure from
string theory or by supersymmetry arguments. One can study
the problem how these terms will affect the classical solution  
\cite{exact}.

\subsection{Stability  of  pp-waves to quantum corrections in  
general covariant
theories}

It is instructive to remind here the situation with the exactness of the
pp-waves in general relativity. Suppose that we have some general  
covariant
theory where the action includes  higher derivative terms/quantum  
corrections
which are general covariant. Pp-wave geometries are space-times  
admitting a
covariantly
constant null vector field as shown by Brinkmann
$
\nabla_{\mu} l_{\nu} = 0\ ,   l^{\nu}l_{\nu}= 0 \ .
$
For instance, for the class of d-dimensional pp-waves
with metrics
of the form  $
ds^2 = 2 du dv + K (u, x^i ) du^2 - dx^i d x^i \ ,
$ the Riemann curvature is \cite{Gu}
\begin{equation}
R_{\mu\nu\rho\sigma} = - 2 l_{[\mu}( \partial_{\nu]}
\partial_{[\rho} K )
l_{\sigma]}\ .
\end{equation}
The Ricci tensor vanishes if $K$ is a harmonic
function in the transverse space:
$
R_{\mu \sigma} = - {1\over 2} ( \partial_{\nu}
\partial^{\nu} K )
 l_{\mu}l_{\sigma}\ ,  R= - {1\over 2} ( \partial_{\nu}
\partial^{\nu} K )
 l_{\mu}l^{\mu}=0
$
The curvature $R_{\mu\nu\rho\sigma}$ is therefore orthogonal to $l^{\mu}$
and
to
$\nabla^
\mu $ in all its indices.
Since $K$ is independent of $v$, the metric solves Einstein equations
$G_{\mu\nu}=0$ if  $\partial^2_T K=0$. Possible corrections to field
equations
may come from
higher dimension operators and depend on the curvature tensors and their
covariant derivatives
\begin{equation}
G_{\mu\nu}= F_{\mu\nu}^{corr} ( R_{\mu\nu\lambda\sigma},
D_{\delta} R_{\mu\nu\lambda\sigma}, \dots)
\label{corrections}\end{equation}
 Corrections to Einstein equations are quadratic or higher order in
curvature
tensors. Note that we do not consider the terms in the r.h.s of eq.
(\ref{corrections}) which vanish when classical field equation are
satisfied. For pp-waves these terms are the Ricci tensor and  
Einstein curvature
scalar
and their covariant derivatives. We construct all possible higher  
order terms
from the Riemann curvature and the covariant derivatives of it,  
which do not
vanish for pp-wave solutions. This serves as an analog of the on-shell
superfields which will be used in supersymmetric theories with maximal
supersymmetry.

Now we may analyse all terms in (\ref{corrections}) depending on Riemann
curvature and the covariant derivatives of it. We find that  there  
is no way to
contract two or more of Riemann
tensors
which will form a two-component tensor to provide the r.h.s. of the
Einstein
equation coming from higher dimensions operators. Therefore all higher
order
corrections vanish for pp-waves solutions \cite{Gu}. They remain exact
solutions of
any
higher order in derivatives general covariant theory. This includes
supergravities and string theory with all possible sigma model and string
loop
corrections to the effective action, as long as these corrections respect
general covariance. Note that supersymmetry played no role in establishing
this
non-renormalization theorem.

\subsection {Maximal supersymmetry}

The importance of having the maximal supersymmetry case when considering
quantum corrections to the supersymmetric branes and black holes is  
in the fact
that for a given dimensions the theory is unique, e. g. we have N=1, d=11
supergravity for M2 and M5 branes and N=2 d=10 IIB supergravity for  
D3 case
\cite{exact}. All fields of these maximal supersymmetry supergravities are
sitting in one multiplet, which includes the graviton, there is no  
coupling
between different multiplets.

In maximally supersymmetric case of the near horizon M2, M5  and  D3  
branes the
theories of d=11, d=10 supergravities can be described in the so-called
on-shell superspace, i.e. in terms of superfields which satisfy the  
classical
equations of motion. It is still possible to construct in each case
the manifestly supersymmetric analog of eq. (\ref{corrections}). The  
right hand
side of this equation will depend on available superfields and their  
covariant
derivatives.

The crucial part of the argument which in pp-wave case was the  
presence of the
null Killing vector, here is the fact of the {\it maximal unbroken
supersymmetry} of the relevant solutions. From this one can derive the
characteristic property of the  {\it vacua of M theory and string theory}:
these vacua, $adS_4\times S^7$ + 4-form for the M2 brane,  
$adS_7\times S^4$ +
dual 4-form for the M5 brane and $adS_5\times S^5$ + 5-form for the  
D3 brane,
can be defined completely by the {\it covariantly constant  
superfields}. In
d=11 the basic superfield of Cremmer-Ferrara-Brink-Howe
\cite{CremFer,BrinkHowe} is $W_{abcd}(X, \Theta)$ and for the near horizon
configurations this superfield is covariantly independent  
\cite{exact} on $X$
and on $\Theta$.
\begin{equation}
D_e W_{abcd}(X, \Theta) =D_{\alpha} W_{abcd}(X, \Theta) =0
\end{equation}
In spherical coordinates of eq. (\ref{spherical}) this superfield is  
actually
$X$- and $\Theta$-independent:
\begin{equation}
{\partial \over \partial X^e}  W_{abcd}(X, \Theta) ={\partial \over  
\partial
\Theta_\alpha}   W_{abcd}(X, \Theta) =0
\end{equation}
and is given by the constant value of the form-field of this  
configuration.
This is a generalization to the superspace of the fixed point  
behavior of the
fields near the horizon, established in the usual space in \cite{FKS}.
In cartesian coordinates of eq. (\ref{cartesian}) the superfield is not
constant but only covariantly constant. In case of IIB supergravity  
\cite{JS}
there are two superfields in  Howe-West superspace \cite{BrinkHowe}  
of this
theory, however, they are not independent as there is only one  
supermultiplet
in this theory. One superfield starts with the dilatino,
$\Lambda_\alpha((X, \Theta)$ and was shown in \cite{exact} to vanish for
$adS_5\times S^5$ + 5-form vacuum. The second superfield, $Z^+_{abcde}(X,
\Theta) $ is supercovariantly constant. Here again in spherical  
coordinates of
eq. (\ref{spherical}) this superfield is actually $X$- and  
$\Theta$-independent
\begin{equation}
{\partial \over \partial X^e}  Z^+_{abcde}(X, \Theta)  ={\partial \over
\partial \Theta_\alpha}   Z^+_{abcde}(X, \Theta) =0 \ .
\end{equation}
and equal to the constant value of the RR 5-form.

Thus the correction to classical equations of motion which do not  
vanish on
classical solutions which we discuss here may depend only on  
non-differentiated
value of the superfields. To show that such contributions are  
absent, one has
to observe that the bosonic equations of motion are given by some  
derivatives
of the fermionic equations of motion since they come out as some higher
components of the fermionic equations. The generic form of  
corrections to the
fermionic equations inevitably has to carry a fermionic index. In  
our bosonic
vacuum  such index may come only from a fermionic derivatives on the
superfields defined above. However such derivatives on the  
superfields of our
vacua vanish. This accomplishes the chain of arguments about the  
exactness of
the supersymmetric vacua of M-theory and string theory.

The $adS*S$  vacua form a {\it fixed point in the superspace, where  
the first
derivatives on the superfields vanishes and the superfields take a fixed,
non-vanishing value}. In the string case it is a value of the RR  
5-form, in
M-theory it is the value of the 4-form and its dual. Note that for   
the trivial
vacua, the flat superspace, all these superfields $W_{abcd}(X, \Theta)$,
$\Lambda_\alpha((X, \Theta)$ and $Z^+_{abcde}(X, \Theta) $  vanish  
everywhere.
For generic supergravity they are functions of $(X, \Theta)$. The brane
solutions interpolate \cite{GT}  between these two types of exact  
vacua, flat
superspace and near horizon superspace, to be described below.

\subsection  {$adS_{2}\times S^{2}$ +2-form near horizon black hole  
case with
smaller  supersymmetry}

In case of smaller, non-maximal  supersymmetry in a given dimension,  
the theory
is not completely defined by dimension and the properties of the  
supergravity
multiplet, including the graviton. Therefore the theories with
non-maximal  supersymmetry are not  unique even before higher  
derivative terms
are taken into account. For example, in d=4 N=2 supergravity there  is a
supergravity multiplet, which includes the graviton, and the matter  
multiplets
without a graviton. These are vector multiplets, including gauge  
fields  and
hypermultiplets. Such theories require the information on the  
prepotential, a
function which defines the coupling of the theory. The choice of  
such function
is not unique and we will see below to which extent this affects the  
issue of
exactness.

In \cite{exact} we looked at d=4 N=2 supergravity without vector or
hypermultiplets (pure supergravity), as a toy model for d=11,10  
theories with
one multiplet. In such case there is only one vector field in the theory,
belonging to the supergravity multiplet, no scalars, only one charge  
$Q=Z=M$
and we have classically the Reissner-Nordstrom black hole. Near the  
horizon the
metric tends to the Bertotti-Robinson  $adS_{2}\times S^{2}$  and  
there is a
covariantly constant 2-form. The only superfield of this theory,  
$W_{a b}$ was
shown in \cite{FK} to be covariantly constant due to enhancement of
supersymmetry near the horizon.
The argument about the absence of quantum corrections to supersymmetric
Bertotti-Robinson configuration in pure d=4 N=2 supergravity was  
based on this
fixed point behavior of the supergravity superfield
\begin{equation}
{\partial \over \partial X^e}  W_{ab}^{ij}(X, \Theta)  ={\partial \over
\partial \Theta_\alpha}   W_{ab}^{ij}(X, \Theta) =0 \ .
\end{equation}
as in cases above.

Quite recently some new results on the stability\footnote{The main  
purpose of
\cite{bernard} was to find the corrections to the black hole entropy in
presence of $R^2$ terms. These corrections are due to Wald's  
redefinition of
the black hole entropy in presence of $R^2$ terms and corrections to the
prepotential which takes care of the second Chern class of the Calabi-Yau
threefold. With these modifications the supergravity corrections to  
the entropy
are found to be in agreement with microscopic calculations of the  
entropy by
Maldacena, Strominger, Witten and Vafa \cite{MSWV}. An important step in
establishing this result in  \cite{bernard} was the derivation of the
$adS_{2}\times S^{2}$ solution in presence of $R^2$ terms in the action.}
 of the $adS_{2}\times S^{2}$ geometry with the 2-forms in presence  
of vector
multiplets  and $R_{abcd}^2$ corrections were obtained \cite{bernard}. One
starts with the supergravity coupled to abelian vector multiplets,   
$X^I$ are
the scalar fields of the vector multiplets, and some chiral  
background field
$\hat A$. The coupling is encoded into a holomorphic function  
$F(X^I, \hat A)$
which is homogeneous of degree two. In this theory the lowest  
components of the
reduced chiral multiplet $W_{ab}^{ij}$ related to Weyl multiplet is  
the tensor
$T_{ab}^{ij}$. The background chiral multiplet $\hat A$ is identified with
$W^2$ at some point. This allows to generate the curvature square terms
$R_{abcd}^2$ in the action in a supersymmetric way. In fact, one  
starts with
the Lagrangian which has a superconformal symmetry, so that the  
action is of
the form
\begin{equation}
16 \pi {\cal L} = - e^{-{\cal K}} R +\dots
\label{Caction}\end{equation}
where
\begin{equation}
 e^{-{\cal K}} = i \left [ \bar X^I F_I(X, \hat A) - \bar F_I(\bar X, \bar
{\hat A})X^I\right]
\end{equation}
If not for the dependence of the prepotential on the chiral field  
$\hat A$,
this would be a Kahler potential of the special geometry. The dots in the
Lagrangian are for the action of the vector multiplets and couplings  
to the
chiral multiplet $\hat A$. The central charge is defined as in the  
usual case
of special geometry, however, the prepotential and all functions of  
it carry
the additional dependence on $\hat A$ (on $R_{abcd}^2$)
\begin{equation}
Z=  e^{{\cal K}/2} (p^I F_I - q_I X^I)
\end{equation}
The superconformal symmetry of the  action (\ref{Caction}) has been  
fixed by
the choice of the gauge
\begin{equation}
 e^{-{\cal K}} = i \left [ \bar X^I F_I(X, \hat A) - \bar F_I(\bar X, \bar
{\hat A})X^I\right] =1
\end{equation}
Note that the presence of $R_{abcd}^2$ terms does affect the choice of the
gauge. In this gauge we have a usual Poincar\'{e} supergravity theory with
supersymmetry, without conformal symmetry.

{}From full supersymmetry at the horizon, in
the presence of the $R^2$ terms, it was found that for nonzero
2-forms (i.e. nonzero electric-magnetic charges) the spacetime remains the
Bertotti-Robinson one:  $adS_2\times S^2$. Furthermore it was
established that $X^I, F_I, \hat A and T^{ij}_{ab}$ are constant. At
this stage it has not yet been shown that there is fixed-point
behaviour. But assuming that the values of the moduli are determined
by the charges, one can invoke symplectic invariance and uniquely
determine  the
relevant equations for the moduli. The metric then equals
\begin{equation}
ds^2= ds^2_{adS} +ds^2_{S} =  -{r^2\over |Z|^2}dt^2+ {|Z|^2\over  
r^2} dr^2 +
|Z|^2 d^2 \Omega
\end{equation}
where the central charge defining the size of the adS throat and the  
radius of
the sphere is related to the 2-form as follows
\begin{equation}
T_{01}^{ij}= -2 \epsilon^{ij} \bar Z^{-1} \ .
\end{equation}
Here the central charge in the chosen gauge at the fixed point is given by
\begin{equation}
Z=   (p^I F_I (X_{fix}, \hat A_{fix}) - q_I X^I_{fix}) \ .
\end{equation}
Subsituting these results into the entropy formula that includes
Wald's modification, one then establishes agreement with the
microscopic entropy as determined in \cite{MSWV}.
Thus one can conclude that for this particular theory of N=2, d=4  
supergravity
with vector multiplets and $R^2$ terms the $adS_{2}\times S^{2}$  
geometry with
the 2-forms defining the size of the radius is a solution.

\subsection{Comment on exactness versus conformal flatness}

In the generic case of M-theory as well as in string theory vacua we  
have not
used the
conformal flatness of the metric to prove the stability of the classical
solution. In fact, near horizon metric of  supersymmetric M2 and M5   
branes is
strictly not conformally flat, as $\omega={p+1\over d-p-3}$ is equal  
to $1/2$
and $2$, respectively. Still in D3 case $\omega=1$ and the metric is
conformally flat. Moreover, the argument in \cite{BG} about  
$R_{abcd}^4$ terms
not affecting the $adS_5\times S^5$ configuration is based completely on
conformal flatness of the near horizon geometry of the D3 brane. The  
resolution
of this puzzle is the following.

The maximal amount of 32 unbroken supersymmetries is valid in  
M-theory as well
as in string theory. The integrability condition of eq. (\ref{kil})  
in both
cases reads
\begin{equation}
\delta \hat \nabla_{[a} \psi_{b]} =  \hat \nabla_{[a} \hat \nabla_{b]}=0
\end{equation}
When translated into the superfield language this allows to prove  
that some
higher in $\Theta$ component of the superfield $W_{abcd}$ in M-theory or
$Z^+_{abcde}$ in string theory, vanishes. In M-theory we get  for the
$\Theta^2$ component of the superfield $W_{abcd}$ (see \cite{CremFer} for
details) the following combination of the Riemann curvatures  $R_{rsmn}$
and 4-form  $F_{tuvw}$
\begin{eqnarray}
 W^{''} \sim {1\over 8} R_{rsmn} \gamma^{mn} + {1\over 2} [T_r^{tuvw},
T^{xyzp}_s ] F_{tuvw} F_{xyzp} + T_{[s}^{tuvw} D_{r]} F_{tuvw}  \ .
\end{eqnarray}
Here $T_r^{tuvw}$ is some combination of $\gamma$-matrices. On the  
near horizon
supersymmetric M2 and M5 brane solutions this expression vanishes,  
i.e. the
term linear in curvature is compensated by terms quadratic in  
form-fields. The
terms with the covariant derivative of the form-field vanish for our vacua
independently of the other terms in this equation.
\begin{eqnarray}
\left[ {1\over 8} R_{rsmn} \gamma^{mn} + {1\over 2} [T_r^{tuvw},  
T^{xyzp}_s ]
F_{tuvw} F_{xyzp}\right]_{vac} =0  \ .
\end{eqnarray}
\begin{eqnarray}
 \left[  T_{[s}^{tuvw} D_{r]} F_{tuvw} \right]_{vac}=0 \ .
\end{eqnarray}

For IIB string theory the second component of the superfield $Z^+_{abcde}$
is also given by some combination of the curvature $R_{abcd}$ and of  
the 5-form
fields of the type \cite{HoweWest}
\begin{equation}
(Z^+_{abcde})^{''} \sim  {1\over 4} (\sigma^{cd})_\gamma^\delta
R_{abcd}
-T_{a\gamma}^{~~\epsilon}T_{b\epsilon}^{~~{\delta}}+
T_{a\gamma}^{~~\bar{\epsilon}}T_{b\bar{\epsilon}}^{~~{\delta}} -D_a
T_{b\gamma}^\delta
\end{equation}
Here the torsion tensor  $T_{b\gamma}^\delta$ is a function of the  
RR 5-form
field. On the near horizon supersymmetric D3 brane solutions this  
expression
vanishes, i.e. the term linear in curvature is compensated by terms  
quadratic
in form-fields \begin{eqnarray}
\left[ {1\over 4} (\sigma^{cd})_\gamma^\delta
R_{abcd}
-T_{a\gamma}^{~~\epsilon}T_{b\epsilon}^{~~{\delta}}+
T_{a\gamma}^{~~\bar{\epsilon}}T_{b\bar{\epsilon}}^{~~{\delta}}\right]_{vac}
=\left[ {1\over 4} (\sigma^{cd})_\gamma^\delta
C_{abcd}
\right]_{vac}  = 0  \ .
\label{weyl}\end{eqnarray}
The   terms with the covariant derivative of the form-field vanish
independently.
\begin{eqnarray}
 \left[  T_{[s}^{tuvw} D_{r]} F_{tuvw} \right]_{vac}=0 \ .
\end{eqnarray}

The important difference with the M-theory case is that the combination of
curvature and forms in eq. (\ref{weyl}) on shell forms exactly the  
Weyl tensor!
The bilinear combination of forms provides the difference between  
the Riemann
tensor and Weyl tensor.
Weyl tensor  vanishes for $adS_5\times S^5$ supersymmetric solution  
and this is
a particular form of the proof of the fact that the superfield   
$Z^+_{abcde}$
is $\Theta$-independent. This particular form of the argument does  
not work in
M-theory, however the fact that a combination of curvature and forms  
vanishes
still works! Thus the unbroken supersymmetry which in all cases  
provides the
$\Theta$-independence of the superfield is the fundamental reason for
exactness. In string case this manifests itself via conformal flatness.

\begin{table}
\begin{center}
\begin{tabular}{|l|l|c|}
\hline
supergravity  brane solution & near horizon metric &  $G/H  $ supercoset
\\\hline\hline
$d=11$ sugra  $M2$ brane $(p=2)$& $adS_{4}\times S^7$ &${OSp(8|4)\over
SO(1,3)\times
SO(7)}$\\
$d=11$ sugra  $M5$ brane $(p=5)$& $adS_{7}\times S^4$&${OSp(6,2|4)\over
SO(1,6)\times
SO(4)}$ \\
$d=10$ IIB sugra   $D3$ brane $(p=3)$ &$adS_{5}\times S^5$&  
${SU(2,2|4)\over
SO(1,4)\times
SO(5)}$\\
$d=6$ $(2,0)$ sugra  self-dual string $(p=1)$ & $adS_{3}\times
S^3$&${SU(1,1|2)^2\over
SO(1,2)\times SO(3)}$\\
$d=4$ $N=2$ sugra  R-N black hole $(p=0)$
& $adS_{2}\times S^2$& ${SU(1,1|2)\over SO(1,1)\times SO(2)}$\\\hline
\end{tabular}
\end{center}
\caption{Supergravity brane solutions with $adS_{p+2}\times S^{d-p-2}$ and
$(p+2)$  form.
\label{tab:sol}}
\end{table}


\section{Flat superspace and near horizon superspace, symmetries of  
extended
objects}
The worlvolume actions of Green-Schwarz superstring and
Bergshoeff-Sezgin-Townsend M2 supermebrane are known in the generic  
background
of supergravity. The coordinates $Z= (X, \Theta)$  of the target  
(super)-space
become functions of the world-volume coordinates of the brane $\sigma^\mu$
\begin{equation}
Z^\Lambda (\sigma) =  (X  (\sigma) , \Theta  (\sigma)  )
\end{equation}
The worldvolume Lagrangians depend on the pullback of the geometric  
objects,
vielbeins and forms,  in the target superspace to the worldvolume

\begin{equation}
{\cal L} [ E_\mu {}^ {\bar \Lambda} = \partial _\mu Z^\Lambda E_\Lambda
{}^{\bar \Lambda} (Z) , A_{\mu_0 \dots \mu_p} = \partial _{\mu_0}  
Z^{\Lambda_0}
\dots  \partial _{\mu_p} Z^{\Lambda_p} A_{\Lambda_0 \dots \Lambda_p} (Z)]
\end{equation}

Thus if we know the supervielbein form
$$E_\Lambda {}^{\bar \Lambda} (Z)$$
and the $p+1$ form
$$A_{\Lambda_0 \dots \Lambda_p} (Z)$$
in the superspace for any supergravity theory, one can use this  
information
to construct the worldvolume actions in any background.

Consider first the {\it flat superspace}. There are no form fields,
\begin{equation}
A_{\Lambda_0 \dots \Lambda_p} (Z)=0
\end{equation}
The supervielbein forms are simple
\begin{equation}
E^\alpha =d\Theta^\alpha \qquad E^a =  dx^a - \bar \Theta \Gamma^a d\Theta
\end{equation}
{\it The isometries of the flat superspace}
\begin{equation}
\delta \Theta =\epsilon \qquad \delta x^a = \bar \epsilon \Gamma^a \Theta
\end{equation}
{\it form the super-Poincar\'{e} algebra. This is the reason why the GS
superstring and BST-supermebrane are `manifestly supersymmetric'.}
For example,  GS classical superstring action depends on the  
pullback to the
world-sheet with coordinates $\sigma^\mu$ of the manifestly supersymmetric
vielbein forms of the flat target superspace:
\begin{equation}
E^\alpha_\mu  \equiv \partial_\mu \Theta^\alpha \qquad E^a_\mu =   
\partial_\mu
x^a - \bar \Theta \Gamma^a \partial_\mu \Theta
\end{equation}
Under the superspace isometries these objects are invariant
\begin{equation}
\delta_{\rm isom} E^\alpha_\mu  =0  \qquad \delta_{\rm isom} E^a_\mu =  0
\end{equation}
and therefore the choice of the background provides the global  
symmetry of the
GS and BST actions.

One would like to construct the string and the M2 and M5 and Dp  
brane actions
not only in the flat superspace but also in the background of the
supersymmetric branes. The most interesting case would be  the IIB string
interaction with RR 5-form of the D3 brane.

To construct the supersymmetric worldvolume actions in any  
background other
than the flat superspace some time ago was looking like an impossible
task. The point is that the vielbeins of M-theory and IIB string  
theory depend
on 32 fermionic coordinates $\Theta$ and therefore they look like
$$E(X, \Theta)_\Lambda {}^{\bar \Lambda}  = (E_0(X))_\Lambda  
{}^{\bar \Lambda}
+ (\Theta E_1(x))_\Lambda {}^{\bar \Lambda} + \dots + (\Theta^{32} E_{32}
(x))_\Lambda {}^{\bar \Lambda}$$
For any particular background one would be able to find such long  
superfield
depending on 32 fermionic coordinates $\Theta$ but one may not  
expect to get
any closed form of it, in general. A beautiful exception from this  
rule is the
superspace generalization of the the near horizon bosonic background  
of M2, M5
and D3 branes (and other cases in the Table 1), suggested in  
\cite{KRR}. The
supercoset construction was developed for the  IIB superstring and  
D3 brane in
\cite{Tsey} and with the use of the closed form of the near horizon  
superspace
\cite{KRR} these actions have been presented in the supersymmetric
$adS_5\times^5$ with RR form  background in a closed form.

One may either use the supercoset construction $G/H$ or  
equivalently, use the
supergravity theory \footnote{See the contribution of P. Claus to these
Proceedings.} to find the near horizon superspace which at $\Theta=0$ is a
bosonic near horizon M2, M5, D3 brane, etc. One starts with the  
superalgebra
${\bf G}$, which for each case is shown in the Table 1. The supercoset
construction $G/H$ consists of solving the set of Maurer-Cartan equations
\begin{equation}
{\cal D}^2=0
\end{equation}
where
\begin{equation}
{\cal D} \equiv d + L^A B_A + L^\alpha F_\alpha
\end{equation}
Here $B$ and  $F$ are bosonic and fermionic generators of the  
superconformal
algebra ${\bf G}$.
The solution of MC equations for fermionic 1-forms takes the  
following form
: there is a term linear in $\Theta$ and higher order corrections  
enter via a
multiple commutators of fermionic generators:
\begin{equation}
F_\alpha L^\alpha = F_\alpha D \Theta^\alpha +
 [ F_\alpha \Theta^\alpha [ F_\beta \Theta^\beta,  F_\gamma D  
\Theta^\gamma ]
+\cdots
\end{equation}
Here the fermionic generator $F$ consists of supersymmetry $Q$ and special
supersymmetry $S$ and
$D $
is the value of the operator ${\cal D}$ at $\Theta=0$ and $L^A_0$  is the
bosonic Cartan form at $\Theta=0$.
The solution for Cartan forms can be written in a closed form as
\begin{eqnarray}
L^\alpha &=& \left(\frac {\sinh {\cal M}}{\cal  
M}D\Theta\right)^\alpha\,,\qquad
L^A = L_0^A + 2 \Theta^\alpha f_{\alpha\beta}^A\left(\frac{\sinh^2
{\cal M}/2}{{\cal M}^2}D \Theta \right)^\beta\,,
\end{eqnarray}
where the matrix ${\cal M}$ is quadratic in $\Theta$ and depends on the
structure constants of the superconformal algebra
\begin{equation}
({\cal M}^2)^\alpha{}_\beta = f^\alpha_{A\gamma} \Theta^\gamma
\Theta^\delta f_{\delta\beta}^A\,.
\end{equation}

The superisometries of this background have been found recently  in  
a closed
form in \cite{CK} and they are given by the transformations of near  
horizon
superspace coordinates $Z$
\begin{eqnarray}
\delta_{adS*S} Z&=&\delta_{adS*S} Z  (Z)
\end{eqnarray}
and the compensating stability $H$-group transformations.
These {\it isometries form a superconformal algebra}. Therefore the  {\it
actions of the extended objects in this background have a  superconformal
symmetry} since the pullback to the worldvolume $\partial _\mu Z^\Lambda
E_\Lambda {}^{\bar \Lambda} (Z)$ of the space-time forms are  
invariant under
the isometries.

The M2 supermebrane classical action in $adS_4\times S^7$ and  
$adS_7\times S^4$
backgrounds of the near horizon M2 and M5 branes has been  
constructed in the
near horizon superspace in \cite{WPPS}.


\section{Issues in gauge-fixing of $\kappa$-symmetry}
The supersymmetric actions of extended objects have local worldvolume
$\kappa$-symmetry in generic background. Therefore   1/2 of the  
32-component
spinor $\Theta$ are unphysical and one have to get rid of them. The  
standard
procedure consists of gauge-fixing of this symmetry, by choosing an  
algebraic
constraint on  $\Theta$ with non-propagating ghosts (if the  
constraint includes
the worldvolume derivatives, one has to consider the ghosts action  
since in
this case the ghost are propagating fields).
In the near horizon superspace one can consider at least 3 possibilities.

\begin{itemize}

\item Light-cone gauge, $\Gamma^+ \Theta=0$ or $\Gamma^- \Theta=0$.

The first possibility is to consider the same gauge which has been  
used for the
quantization of the GS superstring in the flat superspace \cite{GSW}. This
gauge requires the $P^+$ components of the momenta to be  
non-vanishing since
the kinetic term for the remaining fermions  looks as $ \bar \Theta
\Gamma^- P ^+ \partial \Theta$ and one has to be able to divide on $P^+$.
This works well even for massless states for which $P^+P^- +( P^i)^2=0$.
The problem in the near horizon space is that in the light-cone gauge the
values of the vielbein forms do not seem to simplify and each of these
superfields still goes all the way till $\Theta^{16}$. This may not  
necessarily
be a major problem, but still one may try to do different things.  
Note that by
choosing $\Gamma^- \Theta=0$ gauge we would not change anything in proper
notation, it is an equivalent gauge.
\item $Q,S$ class of gauges for the near horizon background \footnote{In
\cite{K} we called these gauges  Killing (anti-Killing) spinor gauge  
for fixing
$\kappa$-symmetry, since the Killing spinors play an important role  
here. On
the other hand there is a choice of the gauge in the superspace  
which was also
given a name of a Killing spinor gauge versus Wess-Zumino gauge  
\cite{KRR}. In
Killing spinor gauge in superspace the spinor-spinor component of  
the vielbein
at vanishing $\Theta$ is taken from the  the Killing spinors of the  
bosonic
space \cite{LPR}. To avoid misunderstanding we will refer
to the relevant gauge in the superspace as to `Killing spinor gauge'  
and to the
gauges  for fixing $\kappa$-symmetry, as to $Q$ or $S$ gauge.
}
{}.
In the context of the superconformal algebra there is a natural  
split of the
fermions into
\begin{equation}
F= \left (\matrix{
Q_{+1/2}\cr
S_{-1/2}\cr
}\right )
\end{equation}
where the supersymmetry generator $Q$ has a conformal weight $+1/2$  
and the
special supersymmetry generator $S$ has a conformal weight $-1/2$.
The coordinates also can be split in analogous way:
\begin{equation}
F\Theta = Q \Theta_Q + S \Theta_S
\end{equation}
Let us now consider these two inequivalent possibilities. The basic  
reason why
these two gauges are inequivalent is due to the triangular nature of the
supervielbein in the Killing spinor gauge in the superspace. In  
these class of
gauges  $\Theta $
are considered to be the functions of $X$ of the form $\Theta (X) =  
{\cal K}
(X)\theta$, or in split form:
\begin{equation}
\left (\matrix{
\Theta_Q(X) \cr
\Theta_S (X) \cr
}\right )= \pmatrix{
{\cal K}_{Q}{}^+(X)  & {\cal K}_{Q}{}^- (X) \cr
0 & {\cal K}_{S}{}^- (X) \cr
}\left (\matrix{
\theta_+\cr
\theta_-\cr
}\right )
\end{equation}
and $\theta$ are $X$-independent coordinates. In such case one can show
\cite{KRR} that
\begin{equation}
D\Theta = {\cal K} (X) d\theta
\end{equation}
\item $Q$-gauge, $\theta_-=0$,  \cite{K}.

This gauge gives the  remarkable simplification of superspace  
vielbeins. Note
that in this gauge
\begin{eqnarray}
\Theta_Q &=& {\cal K}_{Q}{}^+\theta_+  \qquad (D\Theta)_Q  = {\cal  
K}_{Q}{}^+
d\theta_+    \\
\Theta_S&=&0   \hskip 2 cm  (D\Theta)_S  = 0
\end{eqnarray}
Therefore at $\theta_-=0$
\begin{eqnarray}
{\cal M}^{2n} \Theta &=& 0 \ ,    \quad   {\rm at}  \quad  n=1,2,  
\dots   \\
{\cal M}^{2n} D \Theta&=&0 \ ,  \quad   {\rm at}  \quad  n=1,2, \dots
\end{eqnarray}
The vielbein forms are reduced to the following expressions
\begin{eqnarray}
L_Q &=&(D\Theta)_Q\,,\qquad L_S=0 \ , \qquad
L^A = L_0^A + 2 \Theta^\alpha_Q f_{\alpha\beta}^A (D \Theta) ^\beta_Q\,,
\end{eqnarray}
i.e. the vielbeins are quadratic in $\Theta_Q$ at most, like in the flat
superspace. Therefore the actions in the $Q$-gauge are no more complicated
than those in the flat superspace, what concerns the fermions. One has to
specify the conditions when such gauge is admissible and we will  
give examples
of this for the GS string in $adS_5\times S^5$ in \cite{KR,KT}.
In case we consider the action for the extended object in its own  
near horizon
background, e.g. the D3 brane in $adS_5\times S^5$, a special  
requirement has
to be imposed to make the $Q$-gauge admissible. This requirement is  
that the
momenta in directions transverse to the brane are not vanishing.

\item $S$-gauge, $\Theta_Q=0$.

This constraints on spinors was considered in detail in \cite{SPT}  
with respect
to quantization of the D3 brane. Using our set up we can show that
 at $\Theta_Q=0$
\begin{eqnarray}
{\cal M}^{2n} \Theta &=& 0   \quad   {\rm at}  \quad  n=1,2, \dots  \\
{\cal M}^{2} D \Theta&\neq &0 \ , \qquad  {\cal M}^{2n} D \Theta = 0 \ ,
\quad   {\rm at}  \quad n=2,3,\cdots
\end{eqnarray}
Thus in this gauge
the fermionic vielbeins have terms $\Theta^3$ and the bosonic ones  
have up to
$\Theta^4$. The advantage of this gauge that  one can consider the  
actions of
the extended objects in its own near horizon background without  
requiring the
non-vanishing momenta in transverse directions to the brane.

\end{itemize}
In conclusion, the fixing of $\kappa$-symmetry in supersymmetric $adS*S$
backgrounds has been studied in $Q$-gauge, $S$-gauge and light-cone  
gauge. The
vielbeins depend on up to $\Theta^2$, $\Theta^4$ and $\Theta^{16}$,  
in these
gauges, respectively.

An alternative procedure is available for $adS*S$ spaces, it was called in
\cite{Fre,Pes,SPT} an a priory gauge-fixing. It is based on the  
Supersolvable
subalgebra of the superconformal algebra.

\section{IIB Green-Schwarz superstring in $adS_5\times S^5$ and RR-form
background}

Maldacena's conjecture about the duality between the IIB superstring in
$adS_5\times S^5$ and RR-form and N=4 supersymmetric Yang-Mills  
theory is based
on the fact that both theories have the same symmetry forming the  
$SU(2,2|4)$
superalgebra.

The classical superstring action in this background was constructed  
recently
\cite{Tsey,KRR}
in the background whose coordinates $Z=(X, \Theta)$ form an  
${SU(2,2|4)\over
SO(1,4)\times
SO(5)}$ supercoset space.
\begin{eqnarray}
S =-\frac{1}{2}\int d^2\sigma\ \left(\sqrt{-g} \, g^{ij}
 L_i^{\hat a} L_j^{\hat a} +  4 i \eps^{ij}\int_{0}^1 ds\
 L_{is}^{\hat a} \S^{IJ} \bar \Theta^I \Gamma^{\hat a} L_{is}^{J}
 \right)
\ .
\label{action}
\end{eqnarray}
The coupling to RR-form $F$ is included into a term of the form  
$\partial X
\partial X \Theta  \Theta F$.
Here  $\S^{IJ}\equiv$ diag$(1,-1)$,  \ $I,J=1,2$  and  $\hat
a=(a,a')=(0,...,4,5,...,9)$. The invariant 1-forms
$L^I=L^I_{s=1}, \ L^{\hat a}=L^{\hat a}_{s=1}$
are given by
\begin{equation}
L_s^{ I} = \bigg({\sinh \left({s\cal M}\right) \over {\cal
M}}
D\Theta
\bigg)^{I}\ , \  \quad L_s^{\hat a }=e^{\hat a }_{\hat m} (X) dX^{\hat m}
- 4 i \bar
\Theta^I\Gamma^{\hat a}
\bigg({
\sinh^2  \left({\ha s\cal M}\right) \over {\cal M}^2} D\Theta \bigg)^I\ ,
\label{LI}
\end{equation}
where  $X^{\hat m}$ and $ \Theta^I$ are the bosonic and fermionic
superstring coordinates and
\be
({\cal M}^2)^{ IL}=  \epsilon^ {IJ} (-\gamma^{ a} \Theta^{J} \bar
\Theta^L \gamma^{ a} + \gamma^{ a'}
\Theta^{J} \bar \Theta^L \gamma^{ a'} ) + {1\over 2}
\epsilon^{KL} (\gamma^{ab} \Theta^I \bar \Theta^K \gamma^{ab}
-\gamma^{a'b'} \Theta^I \bar \Theta^K \gamma^{a'b'})
\ , \label{msquare}
\ee
\begin{equation}
(D\Theta)^I = \bigg[  d +{1\over 4}(\omega^{ab} \gamma_{ab}  +
\omega^{a'b'}  \gamma_{
a'b'}  ) \bigg]
  \Theta^I  -{1\over 2} i\epsilon^{IJ}(  e^{ a} \g_a  + i e^{a'} \g_{a'}
)\Theta ^J \ . \label{DTheta}
\end{equation}
The Dirac matrices are split in the  `5+5'  way,
$\Gamma^a= \gamma^a \times 1 \times  \s_1, \ \
\Gamma^{a'}= 1 \times \gamma^{a'} \times \s_2,$
where $\s_k$ are Pauli matrices
(see \ci{Tsey} for details on notation).
This classical action has {\it two type of symmetries}.

\begin{itemize}

\item Global $SU(2,2|4)$ symmetries.

The global symmetries are the near horizon superspace isometries found in
\cite{CK} which form the $SU(2,2|4)$ superalgebra. As explained  
above, these
symmetries act on the coordinates of the superspace: $\delta_{adS*S}  
Z$. The
isometries are functions of $Z$ and of the global parameters of the  
$SU(2,2|4)$
superalgebra. Under these transformations the classical
string action is invariant. One may expect the spectrum of states to  
have the
{\it superconformal symmetry}  $SU(2,2|4)$.

\item Local symmetries

The action is invariant under local symmetries, {\it reparametrization and
$\kappa$-symmetry},
whose parameters depend on the worldsheet coordinates $\sigma$. The  
action in
(\ref{action})  is a particular example of the IIB superstring action in a
generic background of supergravity \cite{GHMNT} where the local symmetries
symmetries are given. These  local symmetries have to be gauge-fixed.

 The important property of   $\kappa$-symmetry in the curved background of
supergravity is that the background has to be on shell. This means in
particular that any brane action  known in generic IIB supergravity  
superspace
 can be coupled consistently to  the near horizon superspace of the  
D3 brane.
For example,
one can couple the GS IIB string, D1, D3 and D5 brane to $adS_5\times S^5$
superspace with RR form.

\end{itemize}

By gauge-fixing
\ksym one can
reduce the number of fermions  by 1/2  to match the number of physical
bosonic
and fermionic degrees of freedom.   The gauge-fixing  of \ksym  was
performed
in \cite{KR} developing  the proposal \cite{K} and
 the
action was
found which has  terms at most
 quartic in fermions.\footnote{A similar action
was found  in
\cite{Pes} using  supersolvable
 ({\it Ssolv})  algebra approach \ci{Fre}.  In
 \ci{KR} and \ci{Pes}
 different choices of bosonic (and fermionic)
 coordinates
were used:  Cartesian
and horospherical
 in \cite{KR} and projective coordinates on $S^5$  in
\cite{Pes}. A change of variables has been found in \cite{Pes1}  
which brings
both actions to the same form.
 }
The  special $\k$-symmetry gauge
 using the  projector parallel to D3-brane directions
  allowed to substantially  reduce the  power of fermionic terms in the
action.

Let us review  the  \ksym
 gauge fixing of this action performed in \cite{KR}.
 We shall use the `D3-brane adapted'  or `4+6'
bosonic coordinates $X^{\hat m}
=(x^p,y^t)$
in which the \ads metric  is  split into  the parts
parallel and transverse to the D3-brane directions
(we take the radius parameter   to be $R=1$)
\be
ds^2 = y^2 dx^p dx^p  + {1 \ov y^2} dy^t dy^t \ ,\ \ \ \
\ \ \ \ \   y^2 \equiv y^t y^t \ ,
\la{met}
\ee
where
$ p=0,...,3,\   t= 4,...,9 $. In what follows the
contractions of the indices $p$
is understood  with Minkowski metric and indices $t$  -- with  
Euclidean metric.
The \ksym gauge is fixed   using the `parallel to  D3-brane'
  $\Gamma$-matrix
projector
\be
\Theta_-^I=0 \ , \ \ \ \ \
\Theta_{\pm}^I\equiv  \Ppm^{IJ} \Theta^J \ , \qquad
\Ppm^{IJ}=\frac{1}{2}\left(\delta^{IJ} \pm
\Gamma_{0123}\eps^{IJ}\right)\  , \ \ \ \ \  \Pp \Pm=0
\ .
\la{gau}
\ee
In `5+5' coordinates ($x^a=(x^p,x^4=y)$ and $\xi^{a'}$ coordinates  
on $S^5$)
one finds   that
($\G_{0123} = i \g_4 \times 1 \times 1$, \ $\omega^{p4}=e^p$)
\begin{equation}
(D\Theta)^I  =
 \bigg[ \delta^{IJ} (d +{1\over 4} \omega^{a'b'} \gamma_{a'b'})
  + {1\over 2} \ep^{IJ} (  e^{ a'} \g_{a'}  - i e^4 \g_4 )
+ {1\over 2} e^p  \g_p \g_4 \Pm^{IJ} \bigg] \Theta^J \ .
\label{DTHETA}\end{equation}
Using that
the $S^5$ part  of the covariant derivative
satisfies  $  D_5^{IJ} \equiv   \delta^{IJ}
 (  d +{1\over 4} \omega^{a'b'} \gamma_{a'b'})  + {1\over 2} \ep^{IJ}
 e^{ a'} \g_{a'} = (\L d \L^{-1})^{IJ}, \  (D_5)^2=0$,
where  the spinor matrix $\L^{IJ}=\L^{IJ}(\xi)$ is a function of the $S^5$
coordinates, one finds  that
$(D\Theta_+)^I$ can be
   written as
\begin{equation}
D\Theta_+ = (d - \ha d \log y + \L d \L^{-1}) \Theta^I_+  =
 y^{1/2} \L \ d \  ( y^{-1/2}\L^{-1} \Theta_+)    \ . \la{red}
\end{equation}
Eq. \rf{red}  suggests
to make  the change of the fermionic variable
$\Theta \to \theta$
\be
\Theta^I_+  = y^{1/2}\L^{IJ}(\xi) \ \theta^J_+ \ , \ \ \ \ \ \
\Pm^{IJ}  \theta_+^J =0 \ ,  \ \ \ \ \ \
D\Theta_+^I = y^{1/2} \L d \theta_+^I \ .  \la{cha}
\ee
If we  further   transform  from the coordinates  $(y,\xi^{a'})$
to the 6-d Cartesian coordinates $y^t$ in \rf{met}, $y^t= {y\ov \sqrt{1 +
\xi^2}} (1, \xi^{a'})$,
that would effectively absorb the matrix $\L$
into an $SO(6)$  spinor rotation. In the Cartesian coordinates $y^t=  
y \hat y^t
$
the  6-d  part of the covariant derivative has the form
$D_6^{IJ}=
\delta^{IJ}
 (  d +{1\over 2}  \gamma_{st} \hat y^s d\hat y^t)
+ {1\over 2} \ep^{IJ}
  \g_{t}  (d\hat y^t  + \hat y^t d\log y)$.
This simplification  is suggested
by   the form of the
  Killing spinors  in   \ads space viewed as the
near-horizon D3-brane
background.  In particular, writing the 10-d  covariant
derivative  (including the  Lorentz connection
and 5-form terms)
in the  `4+6' coordinates in \rf{met}
 one learns  that  when
acting on  the constrained spinor
$\Theta_+$
it becomes simply
$D\Theta_+^I = y^{1/2}  d \theta_+^I, \ \
\theta_+^I \equiv y^{-1/2} \Theta_+^I$.

As a result, ${\cal M}^2 D\Theta_+=0$
and the fermionic sector of the  action  reduces  only to terms
 quadratic   and quartic  in $\theta_+^I$.  Using
$\Pm^{IJ}  \theta_+^J =0$ to
eliminate $\theta^2_+$ in  favour of
 \be \theta^1_+ \equiv \vt   \  \ee
one finds that the \ksym gauge-fixed
  string action in $AdS_5
\times S^5$ background \rf{action}
expressed in terms of  the bosonic coordinates  $X^{\hat m}
=(x^p, y^t)$ and
the
{\it single}   $D=10$ Majorana-Weyl spinor  $\vt$
takes the following simple
 form\begin{eqnarray}
S =-\frac{1}{2}\int d^2\sigma\ \biggl[\sqrt{-g} \, g^{ij}&&  
\hspace{-0.7cm}
\left(y^2(\partial_i x^p
- 2 i \bar
\vt \Gamma^{p} \partial_i\vt)(\partial_j x^p - 2 i \bar
\vt \Gamma^{p} \partial_j \vt) +\frac{1}{y^2} \partial_i y^t
\partial_j y^t \right) \nonumber \\ &&
 +\  4 i \eps^{ij} \partial_i y^t \bar \vt \Gamma^t
\partial_j\vt \ \biggr] \ .
\label{SA}
\end{eqnarray}
The $\Theta \Theta  \del X\del X$ terms representing
the {\it coupling to the  RR background}
   present in  the original action \ci{Tsey} in eq. (\ref{action})
are now `hidden'  in the $\bar \vt \del \vt \del  X $ terms
because of    the redefinition  made in \rf{cha}.
We may also  put the action in the first-order
form by  introducing  the `momenta' (Lagrange multipliers) $P^p_i$  
in the D3
brane directions $0,1,2,3$:
\begin{eqnarray}
S =- { 1 \ov 2}
\int d^2\sigma\ \biggl[\sqrt{-g} \, g^{ij}&& \hspace{-0.7cm}
\left(- {1\over y^2} P^p_i P^p_{j}  + 2  P^p_i (\partial_j x^p - 2 i \bar
\vt \Gamma^{p} \partial_j \vt) +  \frac{1}{ y^2} \partial_i y^t
\partial_j y^t \right) \nonumber \\ &&
+ \ 4 i \eps^{ij} \partial_i y^t  \bar \vt \Gamma^t
\partial_j\vt \biggr] \ .
\label{SimpleAction1}
\end{eqnarray}

One can  use the conformal gauge $g^{ij} = \eta^{ij} f (\sigma)$ to  
fix the
reparametrization symmetry which leads to the standard
 $b, c$ ghosts. The gauge-fixed action becomes
\begin{eqnarray}
S =- { 1 \ov 2}
\int d^2\sigma\ \biggl[ \, \eta^{ij}&& \hspace{-0.7cm}
\left(- {1\over y^2} P^p_i P^p_{j}  + 2  P^p_i (\partial_j x^p - 2 i \bar
\vt \Gamma^{p} \partial_j \vt) +  \frac{1}{ y^2} \partial_i y^t
\partial_j y^t \right) \nonumber \\ &&
+ \ 4 i \eps^{ij} \partial_i y^t  \bar \vt \Gamma^t
\partial_j\vt \biggr]  + S_{ghosts}(b, c) \ .
\label{SimpleAction2}
\end{eqnarray}

  In order to  achieve  an   understanding
of the  \ksym gauge choice in
\rf{gau} it is useful to  study
 the issue of {\it invertibility
of the fermionic kinetic operator} in the actions  
\rf{SA},\rf{SimpleAction1}.
In particular, we shall consider the
flat space case obtained by omitting
(or just treating  as constant)  the $y^2 $ and $1/y^2$ factors
 in the action \rf{SimpleAction2}.
In  general,
the constraints  coming from the  equation of motion for the 2-d  metric
can be written  in terms of the vielbein components of
the  `momentum'  $\Pi^{\hat a}_i$
defined by the $g_{ij}$-dependent
part of the action  which does not include the
 WZ term
 ($z,\bar z = \s \pm \tau, \
\s^0\equiv \tau, \ \s^1\equiv \s$)
\begin{equation}
 \Pi_z  \cdot \Pi_{ z}\equiv \Pi_z^p  \Pi_z^p   + \Pi_z^t \Pi_z^t   =
0\ , \ \ \
\ \
\ \   \Pi_{\bar z} \cdot
\Pi_{ \bar z}\equiv\Pi_{\bar z} ^p  \Pi_{\bar z}^p  + \Pi_{\bar z}^t
\Pi_{\bar
z}^t  =0 \  \la{con}
\end{equation}
Dots stand for the fermionic terms
in the constraints and  as  above  the
 indices $p$ are contracted with 4-d
Minkowski metric  and  the indices $t$ --  with   6-d
Euclidean
metric.
In the case of  the action  \rf{SA}\
$\Pi^p_i = y  ( \del_i x^p
 - 2 i \bar \vt \Gamma^{p} \partial_j \vt)
, \
\Pi^t_i =  y^{-1} \del_i y^t
.$

Before \ksym gauge fixing
the quadratic fermionic terms in the flat-space GS action are
\be
 \bar \Theta^1 (\Pi \cdot \Gamma)_z
\partial_{\bar z }  \Theta^1\equiv  \bar \Theta^1 A_1  \Theta^1
\ , \ \ \ \ \ \ \
 \bar \theta^2 (\Pi \cdot \Gamma)_{\bar z }
\partial_{ z }  \Theta^2 \equiv  \bar \Theta^2 A_2 \Theta^2 \ .
\ee
{\it On  the classical equations and constraints we get
$
 (A_1)^2 =  (A_2)^2 =0  ,$
i.e.
the fermionic operator  is degenerate for any  classical
string background}.
As we shall see below,  after the \ksym gauge fixing  as in  \rf{gau}
the degeneracy is removed   provided the background
is constrained  in a certain way. In case of the gauge-fixed action  
\rf{SA} the
background has to be a BPS one so that the gauge is admissible. Let  
us look at
this in more details.

The  quadratic  term in  the fermionic  part of the
gauge-fixed action \rf{SA} is
(we omit
the fermionic terms in $\Pi$)
\be
\bar \vt  \ y \left[(\Pi \cdot \Gamma)_z
\partial_{\bar z} + (\Pi \ast \Gamma)_{\bar z}
\partial_ z\right] \ \vt \ \equiv\  \  \bar \vt \   A \  \vt \ ,
\ee
where we introduced  the notation
\be
\Pi^p_i  \Gamma^p + \Pi^t_i  \Gamma^t =
(\Pi \cdot
\Gamma)_i \ ,  \ \ \ \ \ \ \ \
\Pi^p_i  \Gamma^p   - \Pi^t_i  \Gamma^t =
(\Pi \ast
\Gamma)_i \ .
\ee
Using the equations of motion for $X^{\hat m}$ and
 the constraints  \rf{con},  the square of the kinetic operator $A$ can be
written as
$$
A^2 = y^2  [ (\Pi \cdot \Gamma)_z (\Pi \ast \Gamma)_{\bar z} +
(\Pi \ast
\Gamma)_{\bar z}  (\Pi \cdot \Gamma)_z] \partial_{ z}
\partial_{\bar z} + ...
$$
\begin{equation}
= \ y^2 [ \Pi^p_z \Pi ^{p}_{ \bar z} - \Pi^t_z \Pi ^{t }_{\bar z} ]
\partial_{
z}  \partial_{\bar z} + ... \ ,
\end{equation}
where dots stand for lower-derivative
 $\del y$ dependent terms
which are absent in the   flat space limit  ($y=\const$).
In flat space
 $A^2$
is invertible even on massless ($M_{10}^2 =0 $)
 10-d string states with
$(\Pi \cdot
\Pi)_\tau=0$ and  $(\Pi\cdot \Pi)_\s=0$   if the  $X^{\hat m}$
background
is a BPS one,
\begin{equation}
 (\Pi^p  \Pi^p)_\tau =- M_4^2\ , \ \ \   \qquad  (\Pi^t  \Pi^t)_\tau
= Z^2 \ ,\   \qquad M^2_4 = Z^2  \ ,
\end{equation}
\begin{equation}
A^2 = -2 y^2 M_4^2  \partial_{ z}  \partial_{\bar z} =
-2y^2 Z^2
\partial_{
z}  \partial_{\bar z} \ .
\end{equation}

 To conclude, we have shown that choosing  the
`D3-brane' or `4-d space-time' adapted  \ksym gauge in the \ads
superstring action    one obtains  an action  in which
the fermionic term is  quartic. The  `4+6'  Cartesian
parametrization of the 10-d space  leads to a  substantial
simplification   of the fermionic sector of the theory.
This
should hopefully allow one
to make progress towards  extracting more non-trivial
information  about  the \ads string theory and thus
about  its dual \ci{Malda} -- $N=4$  super Yang-Mills theory.

\section{Superparticle at the boundary of  \ads; analytic harmonic
superspace of N=2 super Yang-Mills theory}
The string action on \ads is not quadratic and therefore as
different from the flat superspace background it is not clear how
to construct the quantum theory. In curved background one can not
have expected to find a simple quadratic action. However one may
try to find some suitable variables in which the theory will
become more useful. We will try to  make the fermionic action
quadratic in a way relevant to super-Yang-Mills theory. We also
would like to use some guide from the Yang-Mills theory. This
guide is an analytic harmonic superspace\footnote{Recently  the  
limit to the
boundary of the $adS_5\times S^5$ superspace   was used in  
\cite{CKR} to derive
the N=2 harmonic superspace. Here we perform a closely related study  
where we
in addition derive the analytic subspace and get the fermionic action
quadratic.}  where the N=2 d=4
Yang-Mills theory can be formulated off-shell \cite{GIKOS}.

As an example of such a possibility we will consider here a {\it
particular approximation to the full IIB string theory on \ads,
starting with the gauge-fixed string action in the form
(\ref{SimpleAction2})}. The approximation consists of

\begin{itemize}

\item Boundary limit $   |y| \rightarrow \infty $

To take this limit we rewrite the action in spherical coordinates
on $S^5$ which include the radius of the sphere, $|y|$ and 5
angles $\phi_1, \dots , \phi_5$. Only the angular part of the
action $(\partial_i y /y)^2$ given by $\partial_i \phi^{m'}
\partial^i \phi^{n'} g_{m'n'}(\phi)$ survives at the boundary.

The WZ term in the form when the derivative hit the fermions has a
term proportional to the $|y|$. To provide the existence of the
limit is sufficient to require that $\vt$ depend only on $\tau$
and do not depend on $\sigma$. This suggests also the next
approximation:

\item Dimensional reduction of the gauge-fixed string action.

Here we extend the independence of fermionic variables from one of
the world sheet directions, suggested by the boundary limit, to
other fields, coordinates of the 4-dimensional boundary of
$adS_5$,  and angles of the sphere.

\end{itemize}

The toy model action is a gauge-fixed dimensionally reduced to one
dimension string action in the limit to the boundary of \ads. It
is plausible that one can get this action by considering a
superparticle in \ads background and gauge fixing the local
symmetries of the superparticle action and taking the limit to the
boundary. The action is

\begin{eqnarray}
S_{superparticle} =  \int d\tau\ \biggl[ \, && \hspace{-0.7cm}
   P_p (\partial x^p - 2
i \bar \vt \Gamma^{p} \partial \vt) +{ 1 \ov 2} \partial \phi^{m'}
\partial \phi^{n'} g_{m'n'}(\phi)+ S_{ghosts}(b, c)\bigg]\nonumber\\
\nonumber\\
 && = S_{adS}^{boundary} + S_{sphere} + S_{ghosts}\ .
\label{toy}
\end{eqnarray}

The part of the action coming from the $S^5$ part of the geometry
is written here in terms of the independent variables, angles of
the $S^5$. Alternatively one can use from the beginning the action
in the form
\begin{equation}
S_{S^5} = { 1 \ov 2} \int d\tau   L^{a'}(\phi) L^{a'}(\phi) = {
tr \ov c} \int d\tau u^{-1} \partial u u^{-1} \partial u =  {
tr\ov c} \int d\tau  \partial u^{-1} \partial u  \ .
\label{sphere}
\end{equation}
where $L^{a'}, a'=1,2,3,4,5$ are the Cartan form on the sphere
$SO(6)/SO(5)$ and $u(\phi)$ are the coset representative on the
sphere, spherical harmonics taking values in $SU(4)$ algebra.

 The
first term in the action $ S_{adS}^{boundary}$ has a manifest d=4,
N=4 global supersymmetry. We may rewrite it using the
decomposition of one d=10 Majorana-Weyl $\vt$ spinor into 4 d=4
two-component spinors\footnote{Our notation here are as in \cite{GIKOS}.}
$\theta_{\alpha i}, \alpha =1,2 \ , i=1,2,3,4$ and their complex
conjugate, $\bar \theta_{\dot \alpha}^i$.
 \begin{equation}
S_{adS}^{boundary}  = \int d\tau
   P_p (\partial x^p - 2
i \bar \vt \Gamma^{p} \partial \vt) = \int d\tau
   P_p (\partial x^p +
i  \partial \theta^i \sigma^p \bar \theta_i - i  \theta^i \sigma^p   
\partial
\bar \theta_i  )
 \ .
\label{ads}
\end{equation}
The total action (\ref{toy})
\begin{eqnarray}
S_{toy} =  \int d\tau\ \biggl[ \, && \hspace{-0.7cm}
   P_p (\partial x^p  +
i  \partial \theta^i \sigma^p \bar \theta_i - i  \theta^i \sigma^p   
\partial
\bar \theta_i ) +
L_{sphere} [u(\phi)] + S_{ghosts}(b, c)\bigg]
\end{eqnarray}
has superconformal symmetry which follows from the isometry of the  
background
and is non-linearly realized after the gauge-fixing. The global
$Q$-supersymmetry of this action is manifest. It is given by the following
transformations
\begin{eqnarray}
\delta \theta_{\alpha i}&=& \epsilon_{\alpha i} \nonumber\\
\delta \bar \theta_{\dot \alpha}^{ i}&=& \epsilon_{\dot \alpha}^{  
i}\nonumber\\
\delta x^p&=& i(\epsilon^i \sigma^p \bar \theta_i - \theta^i \sigma^p \bar
\epsilon_i)\nonumber\\
\delta u(\phi)&=& 0
\end{eqnarray}
These are precisely the N=4 supersymmetry transformations of global
supersymmetry in the central basis \cite{GIKOS}. The fermionic part of the
action is cubic in fields and the action depends on the 16-component  
spinor.

We would like to make the {\it fermionic part of this action quadratic in
fields}.
It is instructive to truncate the action to  N=2 supersymmetric one.  
For this
we have to take
\begin{equation}
\theta_{\alpha 3}= \theta_{\alpha 4} =0 \qquad \phi_ 3=\phi_4=\phi_5=0
\end{equation}
Thus we cut $S^5$ down to $S^2$ and keep only 1/2 of the fermions.
The action of the sphere variables can be written in two possible  
ways: either
in terms of the angles on the sphere or in terms of spherical harmonics.
\begin{equation}
S_{S^2} = { 1 \ov 2} \int d\tau [ ( \partial \phi_1 )^2  +( \sin \phi_1
\partial \phi_2)^2
 =
tr \int d\tau  \partial u^{-1} \partial u  \ .
\end{equation}
Here one can take  harmonics on $S^2$ in the form suggested in  
\cite{GIKOS86}:
\begin{equation}
u  = \pmatrix{
u^-_1 & u_1^+ \cr
u_2^- & u_2^+ \cr
} = \pmatrix{
\cos {\phi_1\over 2} \;  e^{-i {\phi_2\over 2}} & i \sin   
{\phi_1\over 2} \;
e^{-i {\phi_2\over 2}}  \cr
i \sin  {\phi_1\over 2} \;  e^{i {\phi_2\over 2}}  & \cos  
{\phi_1\over 2} \;
e^{i {\phi_2\over 2}}  \cr
}
\end{equation}
Consider the toy model of the string action, the superparticle  
action at the
truncated boundary  of the near horizon D3 brane:
\begin{eqnarray}
S_{toy} =  \int d\tau\ \biggl[
   P_p (\partial x^p  +
i  \partial \theta^i \sigma^p \bar \theta_i - i  \theta^i \sigma^p   
\partial
\bar \theta_i ) +
{ 1 \ov 2} \int d\tau [ ( \partial \phi_1 )^2  +( \sin \phi_1 \partial
\phi_2)^2]
 + S_{ghosts}\bigg]
\label{toyaction}\end{eqnarray}
{\it The symmetries of this action are the symmetries of the N=2  harmonic
superspace in the central basis } which consists of
$
\bigg (Z^M , u^{\pm} _i (\phi) \bigg)
$ where $Z^M= (x^p, \theta_{\alpha i} , \bar \theta^i_{\dot \alpha})$.

The supersymmetries are shown above.
There are two possibilities to formulate the reality condition of  
this action
which precisely fit the known two possibilities to formulate the reality
condition of harmonic superspace.

1. {\it Standard hermitean conjugation}.
$P_p$ and $x^p$ and angles on the sphere $\phi_1$ and $\phi_2$ are  
real and the
chiral spinors $\theta_{\alpha i}$ and $ \bar  \theta _{\dot  
\alpha}^i$ are
conjugate of each other.

\begin{equation}
\overline P_p = P_p \ , \quad \overline  x^p=x^p \ , \quad \overline
\phi_{1,2}= \phi_{1,2} \ , \quad \overline \theta_{\alpha i} = \bar
\theta_{\dot \alpha}^i
\end{equation}
The action is hermitean conjugate, $\overline S =S$.

2. {\it Hermitean conjugation+antipodal map on the sphere}.
In the central basis this second reality condition is the symmetry of the
action and it means that as before  that $P_p$ and $x^p$  are real,  
the chiral
spinors $\theta_{\alpha i}$ and $ \bar  \theta _{\dot \alpha}^i$ are  
conjugate
of each other. The new feature is the antipodal map on $S^2$ which  
means that
each point on the sphere is projected to the antipodal one by the  
following
shift of the angles:
\begin{equation}
\stackrel{*}{\overline \phi_1 }= \pi-\phi_1 \qquad \stackrel{*}{\overline
\phi_2 }= \pi+\phi_2
\end{equation}
In particular the North pole of the sphere become a South pole under the
antipodal map.
The other variables  are neutral under the antipodal map and  
therefore for them
we have, as before
\begin{equation}
\stackrel{*} {\overline P_p} = P_p \ , \quad \stackrel{*} {\overline  
 x^p}=x^p
\ ,  \quad \stackrel{*} {\overline \theta_{\alpha i}} = \bar \theta_{\dot
\alpha}^i
\end{equation}
The action is hermitean+antipodal map conjugate,  
$\stackrel{*}{\overline S}
=S$.

 It is of particular importance that it is the second {\it reality  
condition
under
hermitean+\\
antipodal map conjugation that is the property of the unconstrained  
analytic
superfields in N=2 analytic superspace}.
This is quite different from the usual chiral superspace of N=1  
supersymmetry
where the chiral superfields in the chiral basis are complex.

Now that we have found this reality condition under
hermitean+antipodal map conjugation to be a symmetry of the toy  
model string
action (\ref{toyaction}), we may ask if the knowledge of the analytic
superspace describing the Yang-Mills theory can help us to make the  
fermionic
part of the action quadratic in fields.
The answer is positive.
As suggested by variables in the analytic basis \cite{GIKOS}
$
\bigg( Z^M_A , u^{\pm} _i \bigg)
$ where $Z^M_A=  (x^p_A, \theta^+, \bar \theta^+, \theta^-, \bar  
\theta^-)$,
we have to perform the following change of the variables in our action:
\begin{equation}
\theta_\alpha ^i = u^{+i} \theta_\alpha ^-- u^{-i} \theta^+_\alpha \  
, \qquad
\theta_\alpha^+ = \theta_\alpha^i u_i^+ \ , \qquad \theta_\alpha^- =
\theta_\alpha^i u_i^-
\end{equation}
and
\begin{equation}
x^p_A = x^p - 2i \theta^{(i} \sigma^p \bar \theta^{j)} u_i^+ u_j^-
\end{equation}
The action becomes
\begin{eqnarray}
S_{anal}=  \int d\tau\ \biggl[
   P_p (\partial x^p_A   +
2 i  \partial \theta^+ \sigma^p \bar \theta^- +2  i  \theta^- \sigma^p
\partial \bar \theta^+ ) +
L(\phi_1, \phi_2,  \theta) + S_{ghosts}\bigg]
\label{anal}\end{eqnarray}
The terms in $L(\phi, \theta)$ in addition to the action depending only on
angles of the sphere has now also some terms with derivatives on  
angles which
depend on fermions $\theta$. We will focus here on the first part of  
the action
which has derivatives on fermions. {\it The nice thing happened: the
derivatives hit only  $\theta^+ , \bar \theta^+$  and not  $\theta^-  
, \bar
\theta^-$. We may introduce the new variables now:
\begin{equation}
\Pi^- = 2i \theta^- P_p\sigma^p \ ,\qquad  \bar \Pi^- = 2i P_p  
\sigma^p \bar
\theta^-
\end{equation}
The part of the action which comes from the boundary of $adS$ after  
truncation
becomes a free quadratic action which depends only on $\theta^+$ and $\bar
\theta^+$ and their canonical conjugate variables, which are related to
$\theta^-$ and $\bar \theta^-$.}
\begin{eqnarray}
S_{anal}^{adS}=  \int d\tau\ \biggl[
   P_p \partial x^p_A   +
 \partial \theta^+ \bar \Pi^- + \Pi^-  \partial \bar \theta^+ \bigg]
\label{free}\end{eqnarray}
Our action is still supersymmetric but the supersymmetry is realized  
on the
smaller set of coordinates related to the {\it analytic  subspace} which
includes
$
 \bigg( \zeta^M , u^{\pm} _i\bigg)
$ where $ \zeta^M= (x^p_A, \theta^+, \bar \theta^+)$.
The off-shell Yang-Mills theory is described by the analytic  
superfields which
in the analytic basis depend only on the coordinates of the subspace. The
analytic subspace is real under hermitean+
antipodal map conjugation. Our action in the analytic basis inherits the
reality condition which includes the antipodal map on the sphere from the
original action in the central basis.

As a final remark in this section we would like to stress that the  
action of
the superparticle on \ads space upon truncation to N=2 supersymmetric part
offers new possibilities to link the string theory with Yang-Mills  
theory via
analytic subspace of the harmonic superspace. One can hope to develop the
analogous methods for the untruncated string theory and simplify the  
structure
of the theory.

\section{Black holes and conformal mechanics}
This section is based on recent work in \cite{kaletal,az,GT2}.
One of the deep issues in black hole physics is the existence of the  
 horizon
which prevents the standard quantum mechanical treatment of this system.
On the other hand there is an issue in the conformal mechanics model of
\cite{DFF} known
as the absence of the ground state with $E=0$. The Hamiltonian of  
\cite{DFF} is
\begin{equation}\label{dffham}
H= {p^2\over 2m} + {g\over 2x^2}\,.
\end{equation}
In the black hole interpretation of the model, the classical analog
of an eigenstate of $H$ is an orbit of a timelike Killing vector  
field $k$,
equal to $\partial/\partial t$ in the region outside the horizon, and the
energy
is then the  value of $k^2$. The absence of a ground state of $H$ at  
$E=0$ can
now be interpreted as due to the fact that the orbit of $k$ with  
$k^2=0$ is a
null geodesic generator of the event horizon, which is not covered by the
static
coordinates adapted to $\partial_t$. The procedure used in \cite{DFF} to
cure this problem was to choose a different combination of conserved  
charges as
the Hamiltonian. This corresponds to a different choice of time, one  
for which
the worldlines of static particles pass through the black hole  
horizon instead
of remaining in the exterior spacetime.

Therefore one can believe that the study of conformal quantum  
mechanics has
potential applications to the
quantum mechanics of black holes.
As one can see from the Table 1. in Sec. 3, the near horizon  
geometry of the
supersymmetric d=4 black holes with electric and/or magnetic charge,  
  is the
simplest example of $adS_{p+2}\times S^{d-4-2}$ with the 2-form  
configuration
with $p=0, d=4$ relevant for the superconformal symmetry of the particle
mechanics.

A  surprising connection between black holes and superconformal mechanics
models of Akulov and Pashnev and of  Fubini and Rabinovici \cite{AP}  
was found
in \cite{kaletal}. The main new observation is that {\it one recovers the
supersymmetric conformal mechanics of \cite{AP} in the limit of the  
large black
hole mass $M\rightarrow \infty$ but one also finds a generalization  
of these
superconformal models  for the black holes with the arbitrary mass $M$}.

We start from the extreme RN metric in isotropic coordinates
\begin{equation}
ds^2 = -\left(1+ {M\over \rho}\right)^{-2}dt^2 + \left(1+ {M\over
\rho}\right)^2\big[d\rho^2 + \rho^2 d\Omega^2\big]\,,
\end{equation}
where $d\Omega^2=d\theta^2 + \sin^2\theta \, d\varphi^2$ is the
$SO(3)$-invariant
metric on $S^2$, and $M$ is the black
hole mass, in units for which $G=1$. The near-horizon geometry is
therefore
\begin{equation}
ds^2 = -\left({\rho\over M}\right)^2 dt^2 + \left({M\over \rho}\right)^2
d\rho^2
+M^2d\Omega^2\,,
\end{equation}
which is the Bertotti-Robinson (BR) metric. It can be
characterized as the $SO(1,2)\times SO(3)$ invariant  
conformally-flat metric on
$adS_2\times S^2$. The parameter $M$ may now be interpreted as the  
$S^2$ radius
(which is also proportional to the radius of curvature of the  
$adS_2$ factor).
A
test particle in this near-horizon geometry provides a model of conformal
mechanics in which the $SO(1,2)$ isometry of the background spacetime is
realized as a one-dimensional conformal symmetry. If the particle's  
mass $m$
equals the absolute value of its charge $q$ then this is just the  
$p=0$ case of
the construction of \cite{conf}.

In horospherical coordinates $(t,\phi=\rho/M)$ for $adS_2$, the
4-metric and Maxwell 1-form of the BR solution of Maxwell-Einstein  
theory are
\begin{eqnarray}
ds^2 &=& -\phi^2 dt^2 + {M^2\over \phi^2}\, d\phi^2 + M^2 d\Omega^2\,,
\nonumber\\
A &=& \phi dt\,.
\end{eqnarray}
The metric is singular at $\phi=0$, but this is just a coordinate
singularity and $\phi=0$ is actually a non-singular degenerate  
Killing horizon
of
the timelike Killing vector field $\partial/\partial t$. We now define a
new radial coordinate $r$ by
\begin{equation}
\phi = (2M/r)^2\, .
\end{equation}
The BR metric is then
\begin{equation}
ds^2 = -(2M/r)^4 dt^2 + (2M/r)^2 dr^2 + M^2d\Omega^2 \, .
\end{equation}
Note that the Killing horizon in these coordinates is now at $r=\infty$.

The (static-gauge) Hamiltonian of a particle of mass $m$ and charge
$q$ in this
background is $H=-p_0$ where $p_0$ solves the mass-shell constraint
$(p-qA)^2 + m^2 =0$. This yields
\begin{equation}
H = (2M/r)^2 \big[ \sqrt{m^2 + (r^2p_r^2 + 4L^2)/4M^2} -q\big]\,,
\end{equation}
where $L^2=p_\theta^2 + \sin^{-2}\theta\, p_\varphi^2$, which
becomes minus the
Laplacian upon quantization (with eigenvalues $\ell(\ell+1)$ for integer
$\ell$). We can rewrite this Hamiltonian as
\begin{equation}
H= {p_r^2\over 2f} + {mg\over 2r^2 f}\,,
\end{equation}
where
\begin{equation}
f = {1\over2} \big[ \sqrt{m^2 + (r^2p_r^2 + 4L^2)/4M^2} + q\big]\, ,
\end{equation}
and
\begin{equation}\label{gee}
g= 4M^2(m^2-q^2)/m + 4L^2/m\,.
\end{equation}
This Hamiltonian defines a new model of conformal mechanics  
\cite{kaletal}. The
full
set of generators of the conformal group are
\begin{equation}\label{fullgen}
H= {1\over 2f} p_r^2 + {g\over 2r^2f}\,, \quad K = - {1\over2} fr^2\,,
\quad D={1\over2} rp_r\,,
\label{HKD}\end{equation}
where $K$ generates conformal boosts and $D$
generates
dilatations.
It may be verified that the Poisson brackets of these generators
close to the
algebra of $Sl(2,R)$.

To make contact with previous work on this subject, we restrict to angular
quantum number $\ell$ and consider the limit
\begin{equation}
M\rightarrow \infty\,, \qquad (m-q)\rightarrow 0\,,
\end{equation}
with $M^2(m-q)$ kept fixed. In this limit $f\rightarrow m$, so
\begin{equation}
H= {p_r^2\over 2m} + {g\over 2r^2}\,,
\end{equation}
with
\begin{equation}
g= 8M^2(m-q) + 4\ell(\ell+1)/m\,.
\label{const}\end{equation}
This is the conformal mechanics of \cite{DFF}. For obvious reasons we
shall refer to this as `non-relativistic' conformal mechanics; the \lq
non-relativistic' limit can be thought of as a limit of large black  
hole mass.
When $\ell=0$ an \lq ultra-extreme' $m<q$ particle corresponds to  
negative $g$
and the particle falls to $r=0$, i.e. it is repelled to  
$\phi=\infty$. On the
other hand, a \lq sub-extreme' $m>q$ particle is pushed to  
$r=\infty$, which
corresponds to it falling through the black hole horizon at  
$\phi=0$. The force
vanishes (again when $\ell=0$) for an \lq extreme' $m=q$  particle,  
this being
a
reflection of the exact cancellation of gravitational attraction and
electrostatic repulsion in this case. A static
extreme particle of zero angular momentum follows an orbit of
$\partial/\partial t$, and remains outside the black hole horizon.

A discussion of  supersymmetric versions of the conformal mechanics  
related to
black holes can be found in \cite{kaletal,az,GT2} and this field  
will require
more studies. For example the superconformal mechanics of the particle
approaching the black hole horizon which is expected to have a full
$SU(1,1|2)$ superconformal algebra, has not been constructed so far  
but it will
be  constructed in the near future.

An interesting aspect of the relation between black holes and integrable
Calogero models has been studied in \cite{GT2}. The model has the  
Hamiltonian
\begin{equation}
H= \sum_i {p_i^2\over 2m} + \sum _{i<j} {\lambda^2 \over (q^i - q^j)^2}\,.
\end{equation}
The case of 2-particle Calogero model was considered in detail. It  
has been
found that to have an agreement with the black hole hamiltonian (in  
case of
large black hole mass) one has to find a way to constrain the  
particle orbital
angular momentum  to even integers so that from Calogero model the  
relevant
coupling constant equals $4l(l+1)$ as in (\ref{const}) at $m=q=1$.
This can be achieved by requiring the identification of the antipodal
points\footnote{This antipodal map on $S^2$ does not seem to be   
related to the
one  described in the context of reality of the analytic subspace of N=2
harmonic superspace. It is interesting, however that in both cases this
operation is particular for the sphere which is part of curved  
geometry under
consideration. Antipodal map  has no analog in the flat space.}
on $S^2$ and opens the possibility to consider Calogero models with  
$q=q_1-q_2$
both positive and negative, which may give some insights about the   
exterior
and also interior of the black hole horizon.

It would be very interesting to understand the quantum mechanical  
features of
the new conformal and superconformal models in \cite{kaletal} before  
the limit
to the large black hole mass is taken. In this limit the curvature of the
$adS_{2}\times S^2$ space vanishes and one may loose the important  
properties
of the curved space. However when the mass of the black hole is not  
very large,
we may find interesting quantum mechanical properties starting from the
classical superconformal mechanics described above.

\section{Concluding Comments}
One of the purpose of this lecture was to explain the new concepts and new
approaches to  {\it strong gravity in the framework of  supersymmetry}.
We are trying  to understand new issues in  superstring theory,  
supergravity
and  superconformal field theories. Our current interest in various  
aspects of
$adS*S$ supersymmetric geometry is based on the fact that this near  
horizon
geometry of D3, M2, M5 branes has 32 unbroken supersymmetry  and is  
exact and
the isometries of the relevant superspace form the superconformal algebra.
Therefore
there is some hope that we are learning new connections between  
algebraic and
geometric concepts   which may survive in the fundamental theory unifying
quantum gravity with other interactions.

\end{document}